# Sensitivity optimization of micro-machined thermo-resistive flow-rate sensors on silicon substrates

**Ferdous Shaun[1], Sreyash Sarkar[1], Frederic Marty[1], Patrick Poulichet[1], William Cesar[1,2], Elyes Nefzaoui[1†] and Tarik Bourouina[1]**

[1]Université Paris-Est, ESIEE Paris, ESYCOM EA 2552, Noisy-le-Grand, France
[2]Now with Fluigent Smart Microfluidics, 1, Mail du Professeur Georges Mathé, Villejuif, France

[†]Corresponding author: E-mail: elyes.nefzaoui@esiee.fr

**Abstract**

We report on an optimized micro-machined thermal flow-rate sensor as part of an autonomous multi-parameter sensing device for water network monitoring. The sensor has been optimized under the following constraints: low power consumption and high sensitivity, while employing a large thermal conductivity substrate, namely silicon. The resulting device consists of a platinum resistive heater deposited on a thin silicon pillar ∼ 100 μm high and 5 μm wide in the middle of a nearly 100 μm wide cavity. Operated under the anemometric scheme, the reported sensor shows a larger sensitivity in the velocity range up to 1 m/s compared to different sensors based on similar high conductivity substrates such as bulk silicon or silicon membrane with a power consumption of 44 mW. Obtained performances are assessed with both CFD simulation and experimental characterization.

## 1 Introduction

An accurate measurement of fluid-flow is of paramount importance in different fields of science and technology such as- gas chromatography [1], environmental monitoring [2,3], weather forecasting [4], bio-sensing [5], medical [6] and biomedical applications [7–9], aircraft monitoring [10] and industrial process control [11]. Flow-rate measurement is crucial for water network management as well. Water is a precious natural resource and therefore is a global priority. Consequently, there is a significant need for an effective measurement system for water networks monitoring in order to reduce wastage and to ensure apposite water quality. Measurement of water flow-rate is necessary not only to survey water consumption but also to estimate and even localize the water loss due to leakage of distribution networks. To obtain a complete and monolithic measurement system for water networks, other parameters need to be considered along with the flow-rate such as pressure, temperature and electrical conductivity. Pressure and flow-rate sensors can provide information about the leakage of the distribution system; whereas the conductivity sensor can indicate the overall water ionic contamination level. The temperature of water is also important since the water conductivity depends on temperature as well. It is therefore tempting to co-integrate all the sensors mentioned above on the same chip so as to provide a monolithic multi-parameter sensing solution, which puts a strong constraint on using silicon as a substrate material[*].

Nowadays, several micro-machined flow-rate measurement systems are available to measure the flow-rate in different fluid media [12,13]. The flow-rate range of such systems range from nL/min to L/min [14,15]. A majority of devices reported in the literature are based on thermal operation principles. Thermal flow-rate sensors can be divided into three categories based on their operation principle- anemometric (Hot-wire), calorimetric and time-of-flight (TOF) [16]. The anemometric flow-rate sensor measures the effect of the flowing fluid on a heating resistance, the hot wire for instance. The velocity of the corresponding fluid can be deduced by measuring the amount of heat extracted by the fluid from the hot

---

[*] Part of this work was presented to the 19th International Conference on Solid-State Sensors, Actuators and Microsystems Transducers 2017, Kaohsiung, TAIWAN, June 18-22, 2017

surface of the sensor. The temperature variation of a hot-wire submitted to a fluid flow, indeed depends on the fluid velocity. Measuring this temperature variation enables the calculation of the fluid velocity. An anemometric flow-rate sensor can be operated in one of the following three modes: constant current, constant temperature or constant power [17,18]. A calorimetric flow-rate sensor measures the asymmetry of the temperature profile caused by the fluid flow around a heating resistor. In general, at least two temperature sensors are placed upstream and downstream the heating resistor for this purpose[3,16]. A TOF flow-rate sensor usually measures the transit time of a thermal pulse over a known distance. This kind of sensor involves two or more resistors. One of them is used as a heater; the others are placed downstream and used as temperature sensors. A short thermal pulse generated by the heater travels through the fluid medium and is sensed by the temperature sensors. The pulse travel time from the source to the sensor enables the calculation of the fluid velocity [19,20].

In the present work, we consider anemometric thermal flowrate sensors operated under constant current. Several versions are designed, characterized and compared: first, a flow-rate sensor, fabricated on bulk silicon. This device exhibits a small sensitivity and a large power consumption. This is mainly due to a large conductive heat leakage through the substrate. We have recently shown that the power consumption and sensitivity of a thermal flow-rate sensor are strongly dependent on the used substrate material [21]. For better performances in terms of power consumption and sensitivity, low thermal conductivity substrates are required. Due to the excellent linearity of its temperature-dependent resistivity, platinum is the most used material to fabricate the heating/sensing element of a thermal flow-rate sensor [15,22,23] while sometimes a mixture of platinum and titanium is also used [24]. Most of the micro-machined thermal flow-rate sensors are fabricated on silica glass [25–27] for maximum thermal efficiency and due to its low thermal conductivity. Sometimes, for nanofluidic or biomedical applications [28,29] for example, silicon, which is not the optimal choice from thermal considerations, is used. Hence, when the use of silicon as a substrate material is a constraint, other solutions are required so as to optimize heat transfer in the device. Therefore, modification of the material's geometry is the only option to achieve low thermal conductance and maximize the device sensitivity. First, a bulk silicon substrate device is considered as a reference. Then, a second version of the sensor, which is based on a silicon membrane is inspected, but this version renders unsatisfactory results. Therefore, further optimization is required. A suspended resistor can be considered and have already been suggested by Neda et al. as early as 1995 [30] and as of late by various works [20,31]. However, such geometric configuration increases the device fragility, particularly if wastewater is targeted as a possible operation medium. An intermediate configuration is then proposed in the present paper: the heater resistor is fabricated on a thin silicon micro-wall at the middle of a large cavity etched on bulk silicon. Such configuration would reduce the conduction heat loss while ensuring an acceptable robustness of the heater/sensor structure. The numerical optimization and design, as well as the experimental characterization of this improved micro-machined thermal flowrate sensors, are reported in the following sections. Obtained performances are also compared to those of the previously designed and fabricated devices.

The manuscript is organized as follows: first, modelling and fabrication processes of the different devices are described. Then, the experimental setup and numerical methods used are presented. Finally, the main numerical and experimental results regarding the sensors' responses to the fluid flow rate are reported and discussed. A particular attention is paid to the comparison of the new pillar-based device and previous prototypes, with respect to their sensitivity to the fluid velocity and their power consumption.

## 2 Methods
*2.1 Design and Fabrication*

The multi-parameter sensor has a MEMS chip size 1 x 1 cm in which the lateral size of each sensor is no more than a few millimetres.

The first prototype of the multi-parameter sensing module is presented in Figure 1. The sensor chip is placed on a PCB, which contains the CMOS chip and other electronics for wireless data transmission as shown in Figure 1(b). Finally, it is inserted into a hollow plastic pole (Figure 1(c)) which will be employed in the water distribution network to measure the multiple parameters.

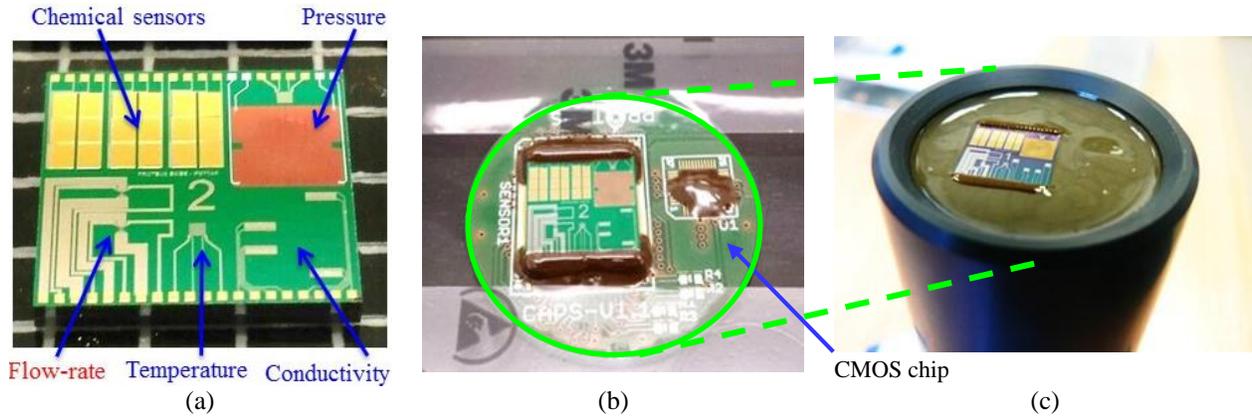

**Figure 1.** Multi-parameter sensor for monitoring water networks- (a) the MEMS chip co-integrating 4 physical sensors (in addition to 5 chemical sensors); (b) capsule board with MEMS and CMOS chip, (c) sensor head.

In Figure 2, the detailed description of different sensing parts of the fabricated chip is illustrated. The chip wire bonding to a PCB, which is designed to build the sensors' electrical connections for lab testing, is shown in Figure 2(a).

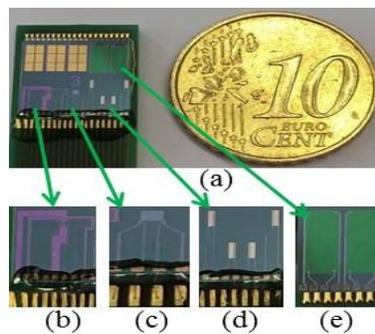

**Figure 2.** Details of (a) fabricated silicon multi-parameter sensor chip including (b) flow-rate sensor, (c) temperature sensor, (d) 2 conductivity sensors and (e) 2 pressure sensors.

The entire PCB used for laboratory characterization is shown in Figure 3. The whole PCB can be divided into 2 parts- base and body. The entire PCB length is 7.2 cm, where the base length is 0.89 cm. The width of the base and the body are 3.82 cm and 1.18 cm respectively. Two sets of header holes are created at the PCB base in order to build the electrical connection between the PCB and the electronic circuit.
A 1 x 1 cm square shaped space is created at the tip of the PCB body to place the sensor chip. An adhesive is used to glue the chip on the PCB. The electrical connection between the chip's connection pads and the PCB's wires is done by gold wire bonding. The electrical connections between the chip and the PCB are protected hermetically by epoxy resin to make it aqua-resistive.

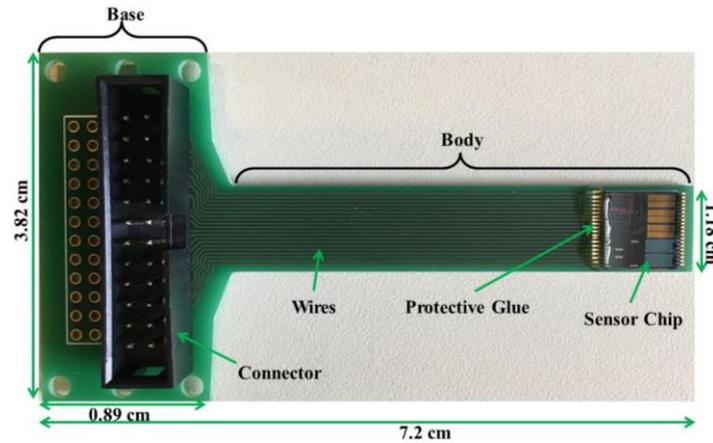

**Figure 3.** PCB for conducting experiments with the sensor chip.

Three different configurations of the flow-rate sensor are considered in the present work. The geometry and composition of the three devices are optimized in order to increase the sensitivity while maintaining low power consumption and satisfactory device robustness. A schematic diagram of the three devices considered is shown in Figure 4. All flow-rate sensors contain the same platinum thermistor.

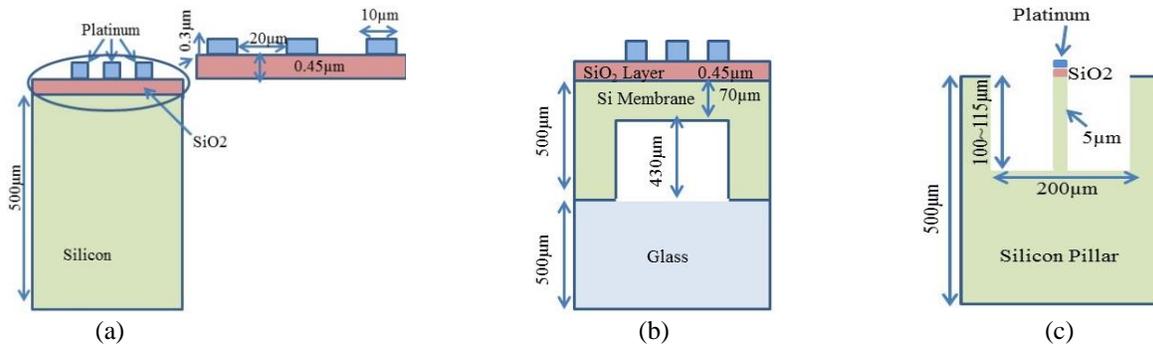

**Figure 4.** The 3 different configurations considered for the flow-rate sensor aiming to study the impact of thermal conductance on the device sensitivity with respect to the velocity- (a) silicon, (b) silicon membrane and (c) silicon pillar structure.

It is noticeable that only the silicon and silicon membrane based sensors contain three platinum resistors while the pillar device contains only one. The arrangement of three platinum resistors allows us to implement any one of the three conventional operation schemes *i.e.,* anemometric, calorimetric and Time-of-Flight (TOF) of a thermal flow-rate sensor. In case of the calorimetric operation principle, the central resistor can be used as a heater and two up and downstream resistors can be used as temperature sensors. Besides, in TOF mode, the left resistor is a thermal pulse generator where the other resistors are temperature sensors depending on the velocity range. Shorter distance between the pulse generator and the sensing resistor exhibits high sensitivity to low velocity, whereas the longer distance enhances the sensor's sensitivity to high velocity [3,32]. The pillar-based device's heater is also made of platinum but with different geometric dimensions. Platinum is chosen for its high TCR (Temperature Coefficient Resistance) and excellent linearity over a wide temperature range. The TCR value of our sputter-deposited platinum is $2.218 \times 10^{-3}$ °C$^{-1}$. Bulk silicon and silicon membrane configurations contain three platinum resistors to enable both calorimetric and TOF operating modes, which have not been considered in this report. The distance between the resistors is 20 μm. The length and width of each platinum resistor are 106 μm and 10 μm respectively, and the thickness is 340 nm. The cavity depth of the pillar-based device is 100~115 μm while the width and length are 100 μm and 200 μm, respectively. The resistor length of this last configuration is equal to the length of the cavity while it is only 5-µm wide. This width is obtained by a parametric numerical optimization of the device sensitivity with respect to its geometric dimensions. Therefore, the resistor width is reduced by half compared to the two other configurations. It can be reduced more, but this would increase the complexity of the fabrication process and makes the resistor more fragile. For the same reason, a fully suspended resistor is not considered. More details on this optimization process are presented in section 3.1.1.

All considered sensors are based on resistive read-out. All of them, except the pressure sensor, are obtained by metal micro-patterning with the combination of titanium, platinum and gold. Further co-integration of the pressure sensor along with the other sensors on the same chip requires additional steps, which include patterning of polysilicon strain gauges and backside etching of the silicon bulk. The latter step is also used to produce thin silicon membranes, not only for the pressure sensor but also for the flow-rate sensor schematically shown in Figure 4(b). The membrane acts either as a mechanically flexible structure or as a thermal insulating layer depending on the considered sensor. A glass substrate is used to support the silicon membrane. The schematic cross-section view of the multi-parameter sensor chip with the silicon membrane flow-rate sensor is presented in Figure 5.

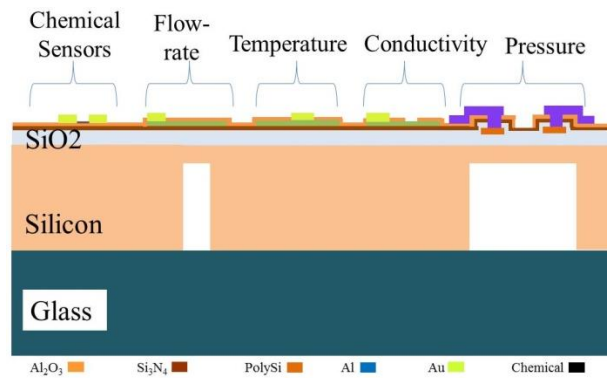

**Figure 5.** Cross-section view of the multi-parameter sensor chip containing the membrane structure flow-rate sensor.

In the case of pillar structure device, a 100 µm deep cavity is created by anisotropic dry etching. The platinum resistor is patterned on the top of the pillar position at the cavity centre. The fabrication sequence is presented in detail in Figure 6(a, b, c, d and e) and Figure 6(f) shows a SEM picture of the pillar device based sensitive element of the sensor which is fabricated in the new version of the multi-parameter chip.

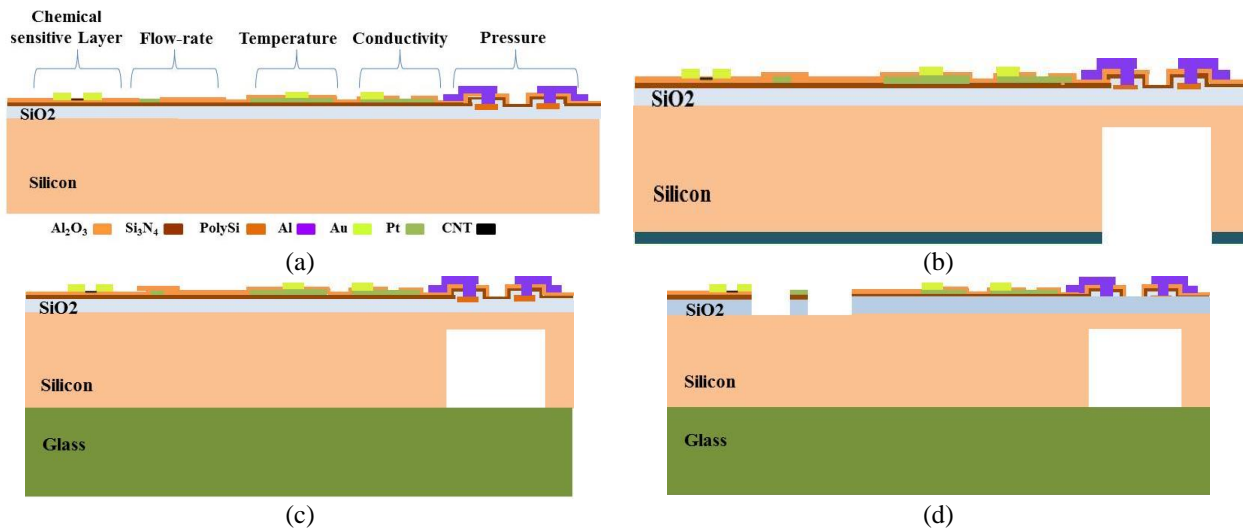

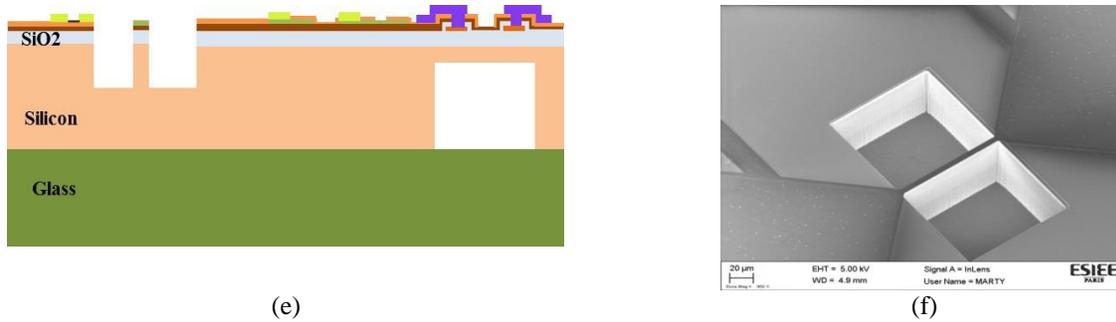

(e)                                                       (f)

**Figure 6.** Schematic of the fabrication sequence- (a) front-side resistors patterning of the multi-parameter sensor chip, (b) patterning the hard mask on the backside for back-side DRIE process in order to create the pressure sensor's membrane, (c) Hard mask removal for silicon-glass anodic bonding, (d) patterning of the front-side in order to create the cavity for the pillar structure flow-rate sensor and etching the front-side deposited layer, (e) the cavity after the front-side DRIE and (f) the SEM photo of the fabricated pillar structure flow-rate sensor.

*2.2 Experimental setup*

The experiment is conducted to extract the sensor's responses with respect to the fluid velocity. The setup is constructed using a rigid pipe with a diameter of 25 mm. The length of the pipe is 10 times its diameter in order to reduce edge effects and to ensure an established flow regime around the sensor. A schematic of the experimental setup is presented in Figure 7. A variable speed pump, submerged in a 20-litre water reservoir, is used to generate water flow at different flow rates. The velocity is measured by following the volumetric measurement process using the same setup (Figure 7). The required time to fill a certain volume of water is measured at different voltage supply levels. Then, the flow rate is calculated. Since the pipe diameter is known, the average velocity is calculated for each flow rate and pump power supply. LabView software interface is used for data acquisition. The sensor is operated according to hot-wire (anemometric) operation scheme under constant current mode. For this purpose, a constant current source is built using an adjustable three-terminal positive-voltage regulator IC device (LM317). To increase the accuracy of measured resistance data, a 4-probe measurement scheme is used so as to eliminate the parasitic wiring resistance effect which may vary due to local heating.

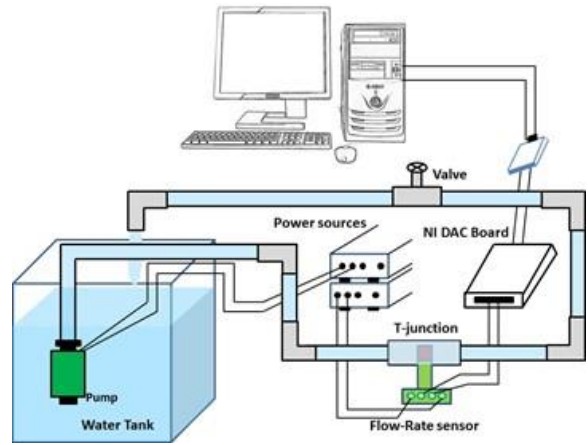

**Figure 7.** Schematic of the experimental setup.

The temperature dependence of the platinum resistance is used to assess the water velocity and flow rate. The setup, indeed measures a change in the resistance value due to a temperature variation induced by the fluid flow. The temperature variation is then obtained using the equation, $R(T) = R(T_0)[(1 + \alpha \Delta T]$; where, $R(T)$ is the resistance value at temperature *T*, $R(T_0)$ the reference value of the resistance at the initial temperature $T_0$, $\Delta T = T - T_0$, is the temperature difference and $\alpha$ is the TCR. The resistance temperature drop with respect to zero velocity situation is then calculated. It is related to the fluid velocity by a calibration process. Consequently, a measurement of the resistance value and hence that of the temperature, enables access to the fluid velocity.

*2.3 Numerical analysis*

The different device configurations presented in Figure 4 are considered for numerical simulations. A parametric study is performed with respect to several parameters such as- shape, geometric dimensions, the substrate material and power consumption. Computational Fluid Dynamics (CFD) simulations are carried out based on Finite Element Method (FEM), using COMSOL Multiphysics. Material properties are imported from COMSOL material library and Conjugate Heat Transfer module is used to account for heat transfer both in the solid and fluid medium. Since the Conjugate Heat Transfer physics is a combination of conductive heat transfer in solid and fluids, and fluid flow (Laminar). Boundary conditions are defined for these two physics. For heat transfer- the left side of the fluidic domain is set to a constant temperature, and a continuity of heat flow is ensured at the other boundaries (Figure 8(a)). The platinum resistor is defined as the heat source with an energy dissipation rate of 10 mW. On the other hand, for the laminar flow physics, the velocity inlet ($v = v_{fluid}$ = constant) and outlet ($\Delta P = 0$) boundary conditions are applied to the left and right side of the water domain respectively, while the top and bottom edges are defined as symmetry axes. All the exterior edges of the sensor body are defined as walls with the 'no-slip condition'. The water medium includes the interior parts of the cavity.

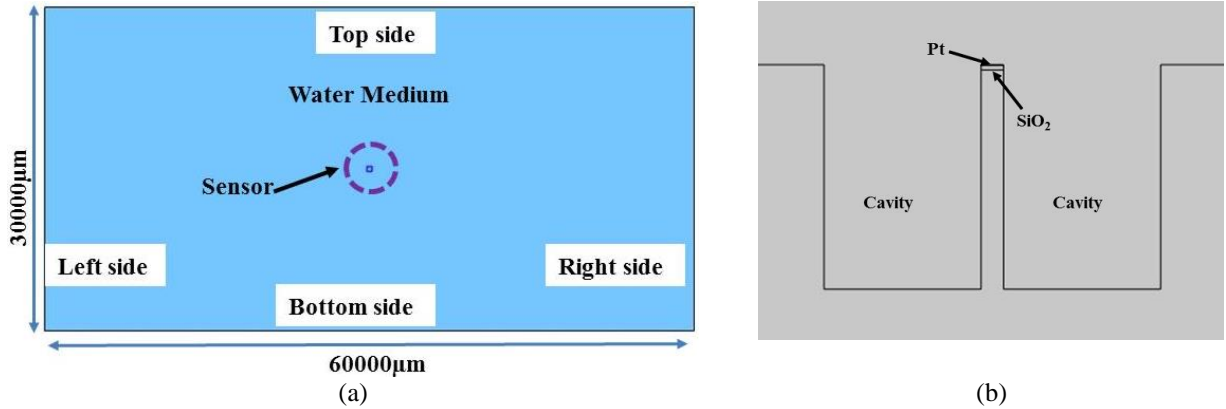

**Figure 8.** Simulation geometry : (a) the whole geometry and (b) zoomed picture of the pillar device

A 450 nm thick silicon dioxide insulation layer is introduced between the substrate and the heater in all configurations. For the membrane sensor, the platinum resistor is mounted on a silicon membrane where the backside cavity depth is around 430 μm. The medium inside the cavity is defined as vacuum. A parametric study is conducted to optimize the aspect ratio of the cavity and the pillar structure. The cavity height and width range from 50 to 300 μm and from 100 to 1000 μm respectively. The pillar height and width are equal to the cavity height and the platinum resistor width, respectively. Because of the almost parabolic velocity profile inside the considered pipes, the sensor is placed on the pipe axis in order to be exposed to the maximum velocity. The sensing element is perpendicular to the main flow direction, thus aligned with the cross-sectional diameter of the pipe. A very large fluidic domain is chosen compared to the sensor size to avoid edge effects.

## 3 Results and Discussion

This section reports obtained numerical and experimental results. The numerical parametric study enables the determination of the optimal sensor design. Then, the optimal device is fabricated, experimentally characterized and compared to previous designs.

## 3.1 Numerical results
### 3.1.1 The sensor's sensitivity to the resistor's width.

As mentioned in section 2.1, a parametric numerical study is performed to optimize the geometry of the sensing element, thereby optimizing the resistor width of the pillar structure flow-rate sensor. We report in this section on the device sensitivity with respect to the heater resistor width $R_w$. The flow-rate sensor's response for a given velocity strongly depends on the heating resistor width. The latter, indeed affects Joule self-heating and the cooling effect of the fluid flow. We show that a decrease in $R_w$ increases the sensor's sensitivity. Simulations are done at 10 mW power supply under a wide velocity range from $10^{-6}$ m/s to 10 m/s. The length and thickness of the resistor are fixed at 100 μm and 340 nm, respectively, while $R_w$ is varied from 1 μm to 10 μm. For a better assessment of the sensors' response with respect to the velocity, the normalized temperature value of the device is plotted as a function of velocity. The normalized temperature is defined as the ratio between ΔT and $\Delta T_{max}$, where ΔT is the temperature difference between the heater resistor and room temperature at a given velocity and $\Delta T_{max}$ is the same temperature difference at zero velocity. The variation of this non-dimensional temperature provides valuable information about the sensor's response to velocity.

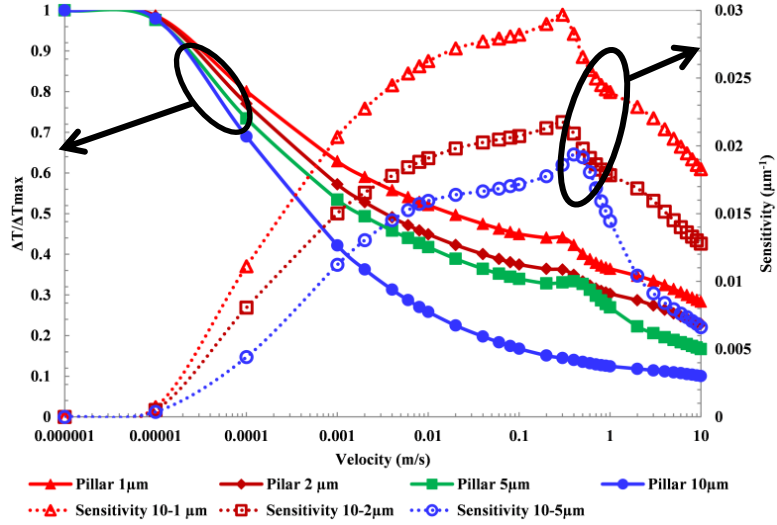

Figure 9. Pillar device normalized temperature and sensitivity plot as a function of the velocity. Here, the primary Y-axis shows the thermal variation, hence the sensitivity of the pillar device with respect to the velocity at different resistor's widths under 10 mW power supply within the velocity range from $10^{-6}$ m/s to 10 m/s. And the secondary Y-axis expresses the sensitivity to the heater resistor width under the same power supply and velocity range.

The normalized temperature (NT) due to water flow is plotted with respect to the primary Y-axis and the NT sensitivity to $R_w$ is plotted with respect to the secondary Y-axis. The effect of the resistor width on the device sensitivity to velocity is evaluated considering the 10 μm width as a reference, where the resultant normalized temperatures for 1 μm, 2 μm and 5 μm wide resistors are compared with the reference resistor response. This sensitivity is calculated as,

$$S_{R_W} = \frac{NT_{R_W} - NT_{R_{Wref}}\big|_v}{\Delta R_W}$$

Where, $S_{R_W}$ is the device sensitivity to the resistor width, $NT_{R_W} - NT_{R_{Wref}}\big|_v$ is the normalized temperature difference between a specific resistor and the reference resistor at a specific $v$; where $v \in [10^{-6}, 10]$ m/s and $\Delta R_W = R_{Wref} - R_W$ is the width difference between a given resistor and the reference resistor.

The best response is observed for the 1 µm wide resistor. On the other hand, the sensor with the broadest resistor (10 µm) rapidly reaches room temperature at high velocity which means a small turn-down ratio compared to the other resistors. In addition to the sensitivity improvement due to $R_w$ reduction observed in Figure 9, we observe a decrease of sensitivity in all cases at high velocities. This is due to the low temperature variation at high velocities since large velocities induce large convective cooling.

In spite of the better performances expected with resistors having smaller widths, the technological fabrication process of 1 µm and 2 µm resistors was found to be significantly more complex. Consequently, the pillar device is fabricated with a 5 µm wide resistor. All experimental results presented and discussed in the following paragraphs were obtained with this device.

*3.1.2   Comparison of the three devices: on bulk silicon, silicon membrane, and silicon pillar*

A numerical study is now conducted to compare the pillar device response, with a 5-µm wide resistor, with the previously fabricated and tested devices, namely the bulk silicon and the silicon membrane device. Simulations are also done at 10 mW power supply under the same velocity range from $10^{-6}$ m/s to 10 m/s. Results are shown in Figure 10, where normalized temperature values are plotted as a function of velocity for the three sensors. For the sake of comparison, NT variations are preferred to absolute temperature values. Bulk silicon and silicon membrane devices exhibit almost similar NT variations with respect to velocity. On the other hand, the pillar structure device shows higher sensitivity within the considered velocity range than the two other devices. This is mainly due to a lower heat leakage to the substrate and a much larger Joule self-heating. The black circle in Figure 10 indicates a discrepancy of the sensor's response for a certain velocity range. A monotonic decreasing temperature is indeed, expected when the flow velocity increases. This discrepancy is due to a reverse flow which appears at this velocity range. The reverse flow velocity field is shown in Figure 11.

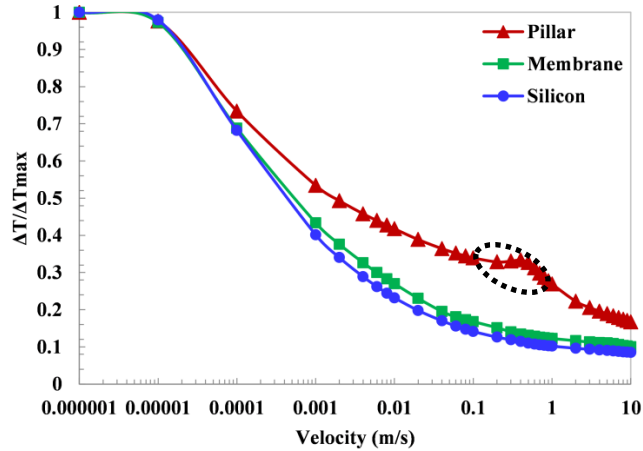

**Figure 10.** Numerical results of the corresponding three sensors response to the velocity at 10 mW.

The main water flow is heated by the heating resistance after a first pass. When the reverse flow carries back this hot water to the heater resistor, the latter experiences a temperature increase. As a result, a velocity increase induces a heater temperature increase rather than the opposite. At 0.3 m/s and 0.4 m/s the pillar flow-rate sensor shows this inconsistent response. A similar reverse flow is observed for bulk silicon and membrane based sensors. However, in these cases, reverse flow fluid is not hot enough to increase the heater temperature because of lower initial Joule self-heating and larger thermal leakage through the substrate. The reverse flow issue at the heater resistor vicinity is fixed by modifying the height and the cavity edge curvature. CFD simulations enable the optimization of these parameters to suppress reverse flow. The re-circulation effect is observed noticeably only for the pillar device among all of the silicon based sensors in the numerical study. The experimental measurement shows a monotonic decrement of the pillar heater temperature within the velocity range from 0 m/s to 0.91 m/s (Figure 14) and the recirculation effect is not observed in practice. The 2D flow assumption in the numerical model may explain the difference between simulation and experimental results where, flow is 3D.

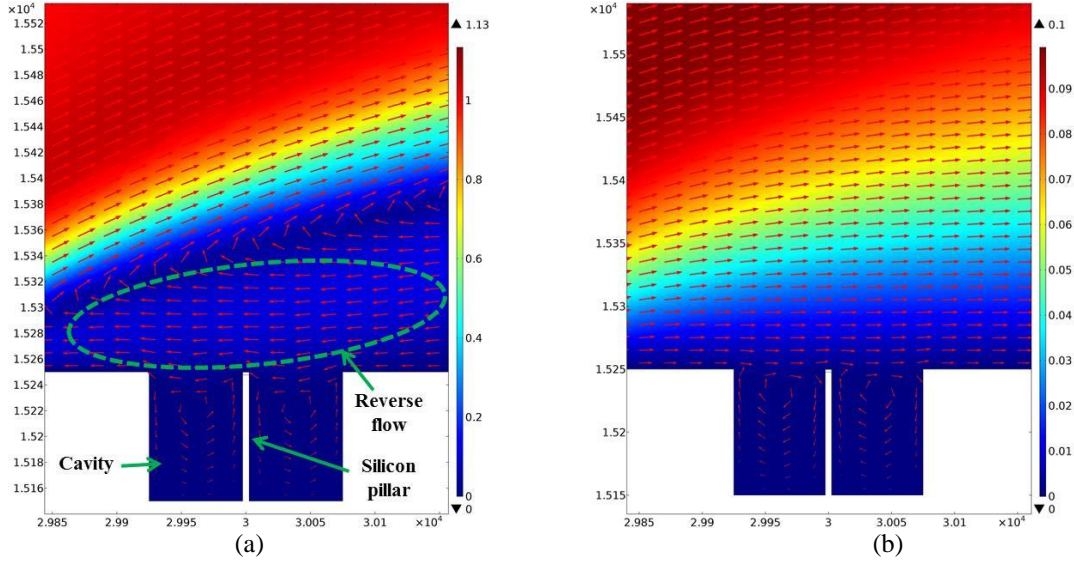

**Figure 11.** (a) Reverse flow at the resistor vicinity illustrating the anomalous behaviour within the velocity range highlighted by black circles in Figure 10; (b) normal velocity profile for other fluid velocities. Only the fluid domain at the heater resistor vicinity is shown here (in blue) while the white domain corresponds to the sensor substrate. The reverse flow is due to the finite size of the sensor chip (not shown here) and the flow separation around it.

*3.2 Experimental results*

The main purpose of this section is to present the experimental responses of the different sensors to the fluid flow. This will be tackled in the last subsection while the two first subsections will tackle static, i.e with no fluid flow, experiments results to discuss the main governing phenomenon for a thermal flow-rate sensor: Joule self-heating.

The resistor's Joule self-heating is an important parameter for a thermal flow-rate sensor. The maximum temperature of the heater resistor under no-flow condition defines the turn-down ratio and the sensitivity of the sensor with respect to velocity. Moreover, the relation between the initial temperature and the power supply indicates the device power efficiency. A large temperature increase at low power consumption leads to a low power consuming thermal flow-rate sensor. Therefore, to determine the considered sensors' efficiency, both Joule self-heating and power consumption tests are studied under zero velocity condition. Results are presented in the following section.

*3.2.1 Joule self-heating and Power consumption.*

Information on Joule self-heating at the resistor level under the no-flow condition helps to anticipate the sensor's response in terms of the velocity and the minimal operating current. The sensor is submerged under the non-flowing water and supply current is varied from 10 mA to 30 mA with a step of 5 mA. Results of Joule self-heating tests are presented in Figure 12(a). We observe that the pillar sensor shows a maximum temperature increase, which is 2.57 °C at 30 mA current supply. Besides, the bulk silicon and the silicon membrane devices show small temperature increase and the maximum temperature values for these two sensors do not surpass 0.8 °C for maximum current supply. Although the three sensors are fabricated on the same substrate material, pillar configuration flow-rate sensor exhibits a significantly larger temperature increase than the others due to an optimized geometry.

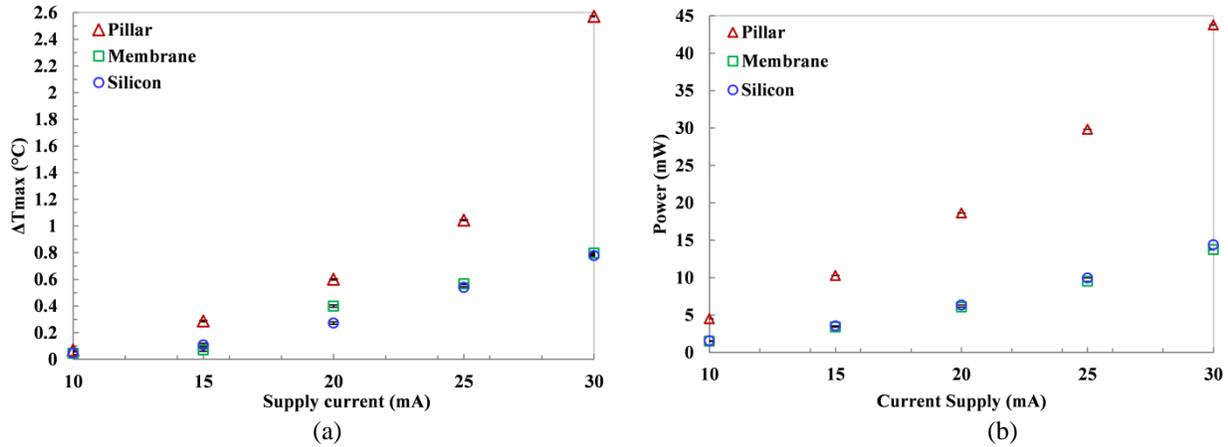

**Figure 12.** (a) Joule self-heating temperature of the heater resistors at different supply current under the no-flow condition, (b) power consumption of the three corresponding sensors.

The power consumption of the three devices is also calculated under the same current supply range. It is illustrated in Figure 12(b). The pillar device consumption is relatively high, up to 43.78 mW for a 30mA current supply. On the other hand, bulk silicon and silicon membrane sensors are three times less power consuming than the pillar configuration at the same current supply. This is due to the high initial resistance value of the pillar device, which is 43.6 Ω. According to Ohm's law, power is proportional to resistance. Therefore, the larger the resistance value, the larger the power consumption. The pillar device heater resistor's width is reduced by two times compared to bulk silicon and membrane resistors, which causes a larger resistance value at room temperature. Consequently, the pillar device is expected to have a power consumption two times larger than the other device. This is not what is observed in Figure 12(b). At the same current supply, a larger resistance value leads to a larger Joule self-heating, hence corresponds to a larger equilibrium temperature. Joule self-heating is then amplified by two phenomena: i) the temperature increase due to Joule heating induces an increase in the resistance values, which results in an additional second-order Joule heating and a temperature increase; ii) The pillar device resistor contact with the substrate due to the pillar structure which reduces heat leakage through the substrate, increases the device equilibrium temperature. This double amplification of Joule self-heating for the pillar device, induces a larger increase of the resistance value, hence a larger power consumption. This amplification can be deduced from the variation of the sensors' resistance values reported in Table 1. For a current supply between 0 and 30 mA, we observe a relative variation of the pillar device resistance of 11.5 % versus a variation of 1 % for the other two devices. The relative resistance variation, hence the Joule self-heating, is larger by one order of magnitude for the pillar device. The tabulated resistance variation values are related to the temperature variation obtained from the Joule self-heating test that is presented in Figure 12(a).

**Table 1.** Resistance variation with respect to the current supply.

| Sensor | Resistance at room temperature (Ω) | Resistance variation range (Ω) |
|---|---|---|
| Pillar | 43.6 | 43.6 ~ 48.65 |
| Silicon membrane | 15.07 | 15.07 ~ 15.27 |
| Bulk silicon | 15.78 | 15.78 ~ 15.97 |

*3.2.2 The sensor's response to the velocity.*

As mentioned above, the velocity is calculated according to the cooling magnitude of an anemometric flow-rate sensor. Therefore, we experimentally measure the sensor's response within a velocity range from 0 ms$^{-1}$ to 0.91 ms$^{-1}$. The flow direction is parallel to the resistor's width. A schematic diagram of the flow direction with respect to the resistor is illustrated in Figure 13.

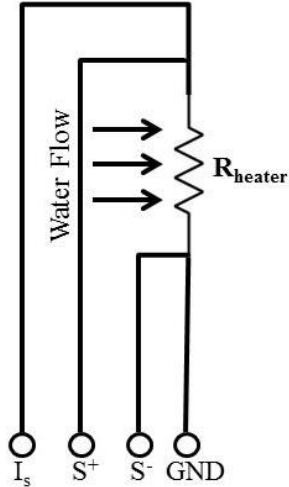

Figure 13. Water flow direction in parallel with respect to the resistor width. Here, I$_s$ denotes supply current, S$^+$ and S$^-$ is the voltage reading ports and GND stands for ground.

We show in Figure 14 the responses of the sensors, *i.e.* the decrement of the normalized temperature, as a function of the fluid velocity. Data are obtained with a 30 mA power supply. For each velocity value, plotted data are the averages of 30 identical response measurements. The resulting experimental errors are also shown in the plot.

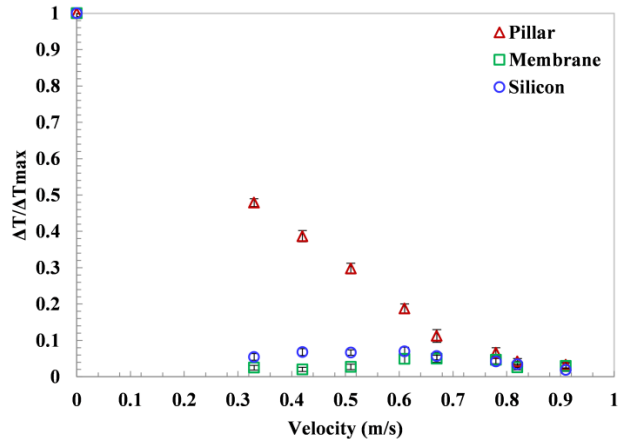

**Figure 14.** Pillar, silicon membrane and bulk silicon flow-rate sensor measured response as a function of velocity at 30 mA current supply.

We observe in Figure 14, a larger sensitivity for the pillar structure sensor as expected from numerical results. The relative temperature variation with respect to the velocity of the pillar flow-rate sensor exhibits a monotonic decrease over almost the entire considered fluid velocity domain, whereas, the silicon bulk and membrane based sensors show almost no temperature variations for velocities larger than 0.3 m/s. The device time constant, defined as the time needed to reach 63% of thermal equilibrium temperature for a given velocity, for pillar, silicon membrane and bulk silicon flow-rate sensors are 1.7 s, 1.9 s and 2.1 s respectively. Experimental results obtained under fluid flow are consistent with preliminary Joule self-heating tests and simulation results.

## Conclusions

We demonstrated that the performance of the thermal flow-rate sensor largely depends on the thermal conductance of the substrate material, which has a great impact on the device sensitivity and power consumption. When there is a limitation regarding the selection of the substrate material, low thermal conductance can be achieved in spite of using high thermal conductivity materials by geometric optimization. In addition, the resistor dimensions are also a governing parameter of the sensor's sensitivity. We demonstrated the successful co-integration of a thermal flow-rate sensor with other MEMS sensors requiring silicon as a substrate material. Unfortunately, the relatively large thermal conductivity of silicon reduces the performance of the thermal flow-rate sensor. Consequently, the main challenge was to achieve a sensitive flow-rate sensor on a silicon substrate. This was successfully achieved through micro-structuration of the silicon substrate using a pillar-like structure and reducing the substrate overall thermal conductance. In the present work, the flow rate sensor is operated independently from the other sensors of the chip. Due to a large number of sensing elements on a very small footprint, cross-interactions between the different sensors is possible and might have an effect on the overall chip performances. The investigation of these cross interactions is the next step of the present project.

## Acknowledgement

Authors would like to thank Hugo Regina for his contributions to the present project. This work received funding from the European Union's H2020 Programme for research, technological development and demonstration under grant agreement No644852. Part of this work received the support of the National Research Agency (ANR) in the frame of the EquipEx project SENSE-CITY of the programme d'Investissements d'Avenir (PIA), involving IFSTTAR and ESIEE Paris as founding members of the consortium. Fabrication of the chips is done in the cleanroom facilities of ESIEE Paris, whose technical staff is deeply acknowledged.